\documentclass[aps,prd,twocolumn,amsmath,amssymb,amsfonts,nofootinbib,superscriptaddress,altaffilletter]{revtex4-1}
\usepackage{graphicx}
\usepackage{hyperref}
\usepackage{verbatim}
\hypersetup{
  bookmarksopen=true
}
\usepackage{amssymb}
\usepackage{amsmath}
\usepackage{color}
\usepackage{supertabular}
\usepackage{dcolumn}

\def\be{\begin{equation}}
\def\ee{\end{equation}}
\def\bea{\begin{eqnarray}}
\def\eea{\end{eqnarray}}

\begin{document}

\title{A semi-coherent analysis method to search for continuous gravitational waves emitted by ultra-light boson clouds around spinning black holes}

\author{S. D'Antonio}
\affiliation{INFN, Sezione di Roma Tor Vergata, I-00133 Roma, Italy}
\author{C. Palomba}
\affiliation{INFN, Sezione di Roma, I-00185 Roma, Italy}
\author{S. Frasca}
\affiliation{INFN, Sezione di Roma, I-00185 Roma, Italy}
\affiliation{University of Rome ???La Sapienza???, I-00185 Roma, Italy}
\author{G. Intini}
\affiliation{INFN, Sezione di Roma, I-00185 Roma, Italy}
\affiliation{University of Rome ???La Sapienza???, I-00185 Roma, Italy}
\author{I. La Rosa}
\affiliation{INFN, Sezione di Roma, I-00185 Roma, Italy}
\author{P. Leaci}
\affiliation{INFN, Sezione di Roma, I-00185 Roma, Italy}
\affiliation{University of Rome ???La Sapienza???, I-00185 Roma, Italy}
\author{S. Mastrogiovanni}
\affiliation{INFN, Sezione di Roma, I-00185 Roma, Italy}
\affiliation{University of Rome ???La Sapienza???, I-00185 Roma, Italy}
\author{A. Miller}
\affiliation{INFN, Sezione di Roma, I-00185 Roma, Italy}
\affiliation{University of Rome ???La Sapienza???, I-00185 Roma, Italy}
\affiliation{University of Florida, Gainesville, FL 32611, USA}
\author{F. Muciaccia}
\affiliation{INFN, Sezione di Roma, I-00185 Roma, Italy}
\author{O. J. Piccinni}
\affiliation{INFN, Sezione di Roma, I-00185 Roma, Italy}
\affiliation{University of Rome ???La Sapienza???, I-00185 Roma, Italy}
\author{ A. Singhal}
\affiliation{INFN, Sezione di Roma, I-00185 Roma, Italy}



\begin{abstract}
As a consequence of superradiant instability induced in Kerr black holes, ultra-light boson clouds can be a source of persistent gravitational waves, potentially detectable by current and future gravitational-wave detectors. These signals have been predicted to be nearly monochromatic, with a small steady frequency increase (spin-up), but given the several assumptions and simplifications done at theoretical level, it is wise to consider, from the data analysis point of view, a broader class of gravitational signals in which the phase (or the frequency) slightly wander in time. Also other types of sources, e.g. neutron stars in which a torque balance equilibrium exists between matter accretion and emission of persistent gravitational waves, would fit in this category. In this paper we present a robust and computationally cheap analysis pipeline devoted to the search of such kind of signals. We provide a full characterization of the method, through both a theoretical sensitivity estimation and through the analysis of syntethic data in which simulated signals have been injected. The search setup for both all-sky searches and higher sensitivity directed searches is discussed.        
\end{abstract}

%
%
\maketitle
\section{Introduction}
\label{sec:introduction}
Among the several possible sources of gravitational waves (GW), those emitting continuous waves (CWs) are one the main targets of current interferometric detectors like LIGO and Virgo \cite{ref:ligo}, \cite{ref:virgo}. Typical predicted sources of CWs are spinning neutron stars (NS) asymmetric with respect to the rotation axis, see e.g.\cite{ref:lasky} for a recent review. Altough no CW detection has been claimed so far, several upper limits have been obtained, which set non-trivial constraints on the source population and characteristics (see e.g. \cite{ref:targ}, \cite{ref:scox1}, \cite{ref:allsky1}, \cite{ref:allsky3}, \cite{ref:narro}, \cite{ref:allsky2} for recent results, and \cite{ref:palo} for a review of the latest results).
Recently, theoretical works have addressed the possibility of CWs emission from a completely different mechanism, namely by ultra-light boson {\it clouds} that would naturally form around black 
holes \cite{arva1}, \cite{brito1}, \cite{arva2}, \cite{brito2}. For stellar mass black holes, and boson masses in the range of ($10^{-14}-10^{-12}$)~\rm eV, the signals would have a frequency in the 
sensitivity band of terrestrial detectors and amplitude such to be potentially detectable if emitted within the Galaxy or, for particulary favourable system 
configurations - high black hole masses and small boson mass - even outside.
The predicted signal is monochromatic (apart from a very small spin-up), but it cannot be excluded that real signals are more complicated and may have a wandering phase (or frequency)\footnote{Further interesting features are, moreover, expected if the cloud is formed around a black hole in a binary system \cite{ref:bhbin}}. While there is still no observational evidence that such mechanism actually takes place, given the increasing sensitivity of 
detectors, it is important to develop a proper data analysis pipeline, which is robust enough to deal with the possible deviation of real signals from the model. Even in case of non-detection, analysis results could be used to 
put interesting constraints on the allowed mass range of ultra-light bosons. 
Moreover, there can be other kinds of sources which emit gravitational-wave signals with similar 
characteristics and for which the chance of detection would be improved by using well designed methods which exploit their specificities. An example is given by neutron stars accreting matter 
from a companion or the sorrounding environment, for which an equilibrium between the angular momentum accreted due to matter infalling on the star's 
surface and the angular momentum loss due to the emission of gravitational waves could be reached, see e.g. \cite{ref:bild1}, \cite{ref:bild2}, \cite{ref:charaba}, \cite{ref:patru}. In this situation the spin frequency of the 
star, and then the signal frequency, would fluctuate around some values as the rate of matter accretion randomly varies in time. 

In this paper we present a robust and computationally cheap method to perform an all-sky, nearly constant frequency, semi-coherent analysis relying on various sets of ``short'' FFTs, each built from chunks of data of different length \cite{ref:pia_sfdb}. 
See e.g. \cite{ref:derga}, \cite{ref:lilli}, \cite{ref:whelan} for different algorithms, developed in the context of CW searches, which try to take into account possible unpredicted phase or frequency variations of the signal, but mainly meant for candidate follow-up or searches toward well localized sky regions.  

The paper structure is as follows. In Sec. \ref{sec:emission} we briefly describe the GW signals we are considering, focusing on those expected from black hole - boson cloud systems and define the parameter space we want to explore. In Sec. \ref{sec:pipeline} we present the analysis pipeline in detail, discussing a possible setup for both an all-sky search with relatively short FFTs and a directed search towards some potentially interesting regions of the sky, like the galactic center, using longer FFTs. Section \ref{sec:sensitivity} is dedicated to the theoretical sensitivity estimation of the method. In Sec. \ref{sec:efficiency} we present validation tests we have done by using simulated signals added to syntethic data. Conclusions are given in Sec. \ref{sec:conclusion}.  

\section{Gravitational wave emission mechanisms}
\label{sec:emission}

The prototype emitter of CW is constituted by a spinning NS, asymmetric with respect to its rotation axis. Many neutron stars, especially those belonging to the class of radio pulsars, are known to be very stable rotators \cite{ref:handbook}, which means that accurate measures of the position and rotational parameters are available. This implies that full-coherent {\it targeted} CW searches over long observation periods (of months or even years) can be done. When the source parameters are uncertain or not known, searches over a wider portion of the parameter space are typically done by using semi-coherent methods. {\it Directed} searches assume the source position is well known, but rotational parameters are largely unknown, while {\it all-sky} searches are carried for sources with no electromagnetic counterpart \cite{ref:palo}. For ``standard'' spinning NSs the signal frequency is expected to slowly decrease over time, due to rotational energy loss in the form of electromagnetic or gravitational radiation. As discussed in the previous section, however, in some cases the emitted signal could be nearly monochromatic, for instance, in the case of a neutron star accreting matter from a companion star, and characterized by random fluctuations of the frequency. 

Recent works \cite{arva1}, \cite{brito1}, \cite{arva2}, \cite{brito2} on the formation of light scalar boson clouds around black holes, and the following GW emission, have provided a new possible interesting source of monochromatic or nearly-monochromatic CWs. The development of the analysis method we present in this paper has been triggered by these theoretical results and is optimized for an all-sky, wide frequency band search of nearly monochromatic signals with possible small frequency fluctuations.
The rest of this section is dedicated to a brief summary of the main features of the GW emission from the aforementioned boson clouds.
Light bosonic fields have been suggested as a possible component of dark matter \cite{ref:arva0}, \cite{ref:marsh}, \cite{ref:hui}. A bosonic field around a rotating black hole can activate a process of superradiant instability \cite{ref:super} in which the field is amplified at the expense of the black hole rotational energy, allowing the formation of a bosonic cloud around the black hole itself. While the equations we present in this section mainly refer to scalar boson fields, recently also the case of superradiance induced by vector fields (such as massive photons) have been addressed at theoretical level \cite{ref:masha}, \cite{ref:cardo}. The instability is strongest if the Compton wavelength of the boson, $\hbar c/m_{\rm b}$, being $m_{\rm b}$ the boson mass-energy in Joule, is comparable to the black hole Schwarzschild radius $2GM_{\rm bh}/c^2$, where $M_{\rm bh}$ is the black hole mass. The instability stops at the saturation, when the real part of the boson field angular frequency,
\be
\omega_{l,n}=\frac{c^3}{G}\mu \left[1-\frac{1}{2}\left(\frac{M_{\rm bh}\mu}{n+l+1}\right)^2\right]
\label{fln}
\ee
(being $n=0,1,2,...$ the overtone number and $l$ is the spherical harmonic index), is $\lesssim m\Omega_{rm H}$, where $\Omega_{rm H}$ is the black hole horizon angular frequency and $m$, with $|m|\le l$, is the azimuthal number. Eq. \ref{fln} is found using a perturbative approach, which holds if $\mu M_{\rm bh}\ll 1$, where 
\be
\mu=\frac{G}{\hbar c^3}m_{\rm b}
\label{mudef}
\ee
Once the saturation is reached, the boson cloud, which possesses a non-zero quadrupole moment, evolves by emitting long lasting GW.
In the following we report the main relations related to the GW emission by the boson cloud by using standard SI units (instead of the geometrical units used e.g. in \cite{arva1},\cite{brito1}).
The strongest GW emission happens for the mode $n=0,~l=m=1$ and has a frequency
\be
f_{GW}=\omega_{1,0}/\pi=\frac{c^3}{\pi G}\mu \left[1-\frac{1}{8}\left(M_{\rm bh}\mu\right)^2\right]
\label{f_gw}
\ee
By imposing that $f_{GW}$ cannot be larger than two times the black hole maximum horizon spin frequency, i.e.
\be
f_{GW}\lesssim \frac{\Omega_{\rm H,{\rm max}}}{\pi}=\frac{c^3}{2\pi GM_{\rm bh}},
\label{f_gw_forbid}
\ee
a constraint on the allowed boson and black hole masses can be derived. 
In Fig. 1 we show the expected signal frequency $f_{GW}$, computed through Eq. \ref{f_gw}, as a function of the boson mass for a few values of the black hole mass $M_{\rm bh}$ between 5$M_{\odot}$ and 100$M_{\odot}$. 
\begin{figure}[htbp]
\includegraphics[width=10cm]{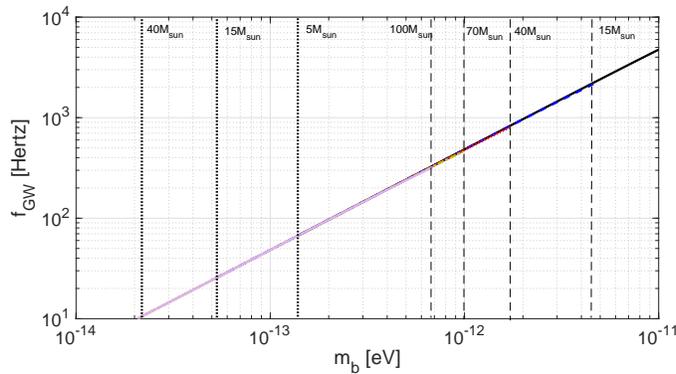} 
\label{fig:mbos_fgw}
\caption{Expected CW signal frequency for boson clouds around black holes, as a function of the boson masses. Different curves identify different values of the black hole mass: $M_{\rm bh}$: $5,15,40,70,100~M_{\odot}$. In fact the curves are nearly superimposed and cannot be easily distinguished. For each black hole mass, the vertical dotted line indicates the minimum allowed boson mass such that the instability grows in a time shorter than the age of the Universe, see Eq.(\ref{tau_inst}) with $\chi=0.5$; the dashed line indicates the maximum boson mass above which the condition of Eq.\ref{f_gw_forbid} is no more fulfilled thus forbidding the presence of heavier bosons.}
\end{figure}  
The signal frequency increases almost linearly with the boson mass, independently of the black hole mass: this happens because the second term in the right hand side of Eq.\ref{f_gw} is much smaller with respect to the first under the assumption $\mu M_{\rm bh}\ll 1$. For each black hole mass, there is an allowed range of boson masses, determined by the condition that the instability develops in a time shorter than the age of the Universe, see Eqs.(\ref{tau_inst}),(\ref{eq:mb_forbid}), and that the condition of Eq.\ref{f_gw_forbid} is verified. The vertical dotted and dashed lines in the figure indicate, depending on the mass of the black hole, such minimum and maximum boson mass\footnote{A possible indirect further constraint, which tends to disadvantage the boson mass range $m_{\rm b}\simeq  [6\cdot 10^{-13},10^{-11}]$ \rm eV, comes from the measurements of black hole angular momentum, see discussion in \cite{arva1}, \cite{ref:cardo}.}. 
The time scale for GW emission is given by \cite{ref:brito3}
\be
\tau_{GW}\simeq 600\chi^{-1}\left(\frac{M_{\rm bh}}{40M_{\odot}} \right)^{-14}\left(\frac{m_{\rm b}}{5\cdot 10^{-13} \rm eV} \right)^{-15}~{\rm yr},
\label{tau_gw}
\ee
where $\chi \in(0,1)$ is the adimensional black hole spin.
This is typically much longer than the instability time scale, in which the cloud grows and saturates, obtained from the imaginary part of the frequency:
\be
\tau_{\rm inst}\simeq 0.008\chi^{-1}\left(\frac{M_{\rm bh}}{40M_{\odot}} \right)^{-8}\left(\frac{m_{\rm b}}{5\cdot 10^{-13} \rm eV} \right)^{-9}~{\rm yr}
\label{tau_inst}
\ee
As stated in \cite{brito1}, both relations are valid in the limit $\mu M_{\rm bh}\ll 1$ and $\chi \ll 1$ (of course, $\chi >0$ in order to have superradiance), although they are still a good approximation when $\mu M_{\rm bh}\sim 1$ and $\chi \sim 1$.
Note the very steep dependence of both $\tau_{GW}$ and $\tau_{\rm inst}$ on $M_{\rm bh}$ and $m_{\rm b}$, which implies that also slightly different values of these parameters can produce large variations on the timescales. In Fig.2 the expected time scale for CW emission, computed using Eq.\ref{tau_gw}, is shown as a function of the boson mass $m_{\rm b}$ for different black hole masses. 
The emission time scale is typically much longer than detector observation time, especially for lower frequency signals for which it can largely exceed the age of the Universe. On the other hand, for very small boson masses, especially in the case of relatively light black holes, the instability time scale can exceed itself the age of the Universe, meaning that GW emission is sub-dominant. More specifically, it is easy to see that $\tau_{\rm inst}>1.3\cdot 10^{10}$ years for masses such that
\be
\left(\frac{m_{\rm b}}{10^{-13}~\rm eV}\right)\le 0.752\left(\frac{M_{\rm bh}}{10M_{\odot}}\right)^{-8/9}\chi^{-1/9}
\label{eq:mb_forbid}
\ee
\begin{figure}[htbp]
\includegraphics[width=10cm]{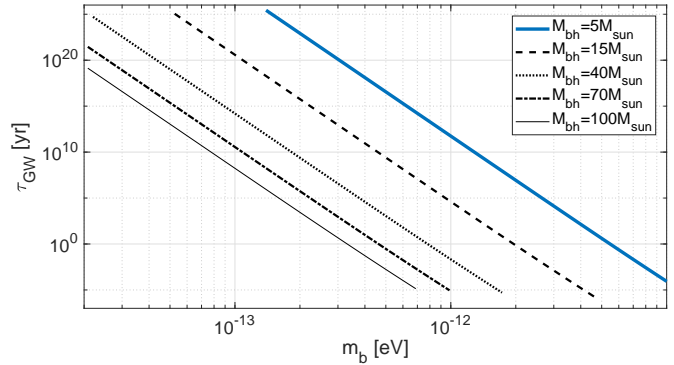} 
\label{fig:mbos_taugw}
\caption{Expected timescale for CW emission, $\tau_{GW}$ (in years), as a function of the boson mass for different black hole masses. For each black hole mass the maximum allowed boson mass is given Eq.\ref{f_gw_forbid}., while the minimum allowed mass is given by Eq. \ref{eq:mb_forbid} with $\chi=0.5$.}
\end{figure}  
The cloud self-gravity causes a very small spin-up of the frequency given, in the case of axion-like particles, by \cite{arva1}
\be
\dot{f}_{GW}\approx 10^{-16}\left(\frac{f_{GW}}{100~Hz}\right)\left(\frac{10^6~{\rm yr}}{\tau_{GW}}\right)\left(\frac{2\cdot 10^{18}~\rm
Gev}{\rm F_a}\right)^2~\frac{Hz}{s}
\label{eq:f_gw_dot}
\ee
where $\rm F_{a}$ is the axion decay constant. 
As it will be clear in Sec. \ref{sec:pipeline}, for most of the accessible parameter space the signal spin-up is negligible and largely within one single bin of the typical search grid.
As GWs are emitted the boson cloud mass decreases according to \cite{brito1}
\be
M_{\rm cloud}(t)=\frac{M_{0}}{1+\frac{t}{\tau_{GW}}},
\label{Mcloud}
\ee
where $M_{0}(M_{\rm bh},m_{\rm b},\chi)$ is the initial cloud mass, whose maximum value is
\be
M_{\rm max}=\frac{G}{c^3}M_{\rm bh}^2\omega_{10}
\label{Mmax}
\ee
The mass cloud is a factor $\sim \mu M_{\rm bh}$ smaller than the black hole mass.
The amplitude of the GW signal, averaged over the source sky position, can be expressed as
\begin{widetext}
\be
h_0(t)\approx 9\cdot 10^{-24}\left(\frac{M_0}{M_{\odot}}\right)^{1/2}\left(\frac{M_{\rm bh}}{40M_{\odot}}\right)^{7}\left(\frac{m_{\rm b}}{5\cdot 10^{-13}~\rm eV}\right)^{13/2}\chi^{1/2}\left(\frac{d}{10~\rm kpc}\right)^{-1}\left(1+\frac{t}{\tau_{GW}}\right)^{-1},
\label{h0approx}
\ee
\end{widetext}
being $d$ the source distance.
Overall, most of the parameter space accessible to current, and next generation, Earth-based inteferometric detectors corresponds to signal frequency below a few hundreds Hertz. More specifically, we take 630 Hz as the maximum frequency in our standard search setup. All the following plots and tables assume the search is done up to this value. Extending the search to higher frequencies would imply an higher computational cost, but not major modifications to the pipeline (see the discussion at the end of Sec. \ref{sec:anasche}). 
    
\section{Pipeline description}
\label{sec:pipeline}
In this section we provide a detailed description of the search pipeline, including the general data organization for each detector, the peakmap database, the steps used to identify candidates and measure their statistical significance, the coincidence analysis among candidates of different detectors and the follow-up.
The basic idea is simple and motivated by the interest to carry out a quick and robust all-sky search for a wide class of CW signals with a small but unpredictable phase or frequency variations during the observation time. 
Some indications on how the pipeline can be adapted to perform {\it directed} searches toward some specific area in the sky, like the Galactic center, will be given in Sec. \ref{sec:direc}.

\begin{figure*}[htbp]
\includegraphics[width=12cm]{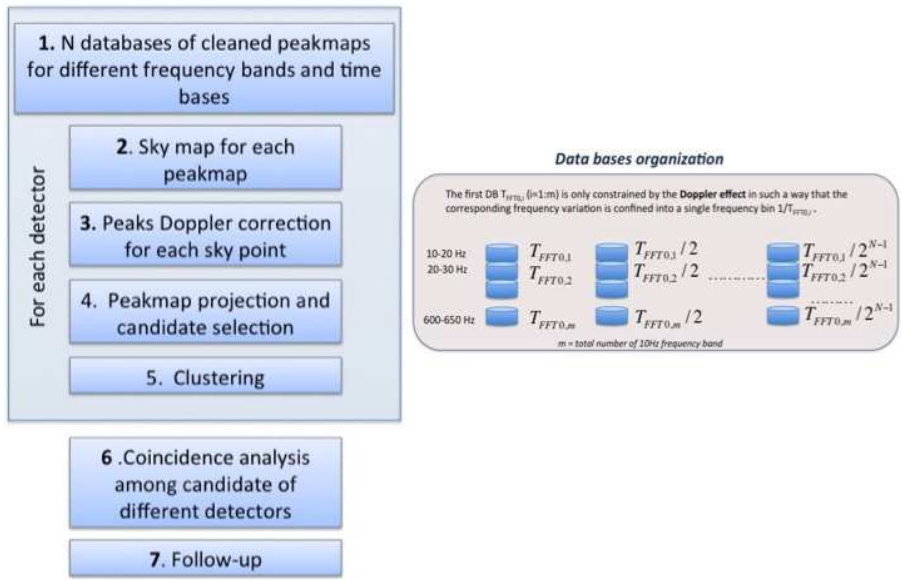} 
\caption{Scheme of the analysis pipeline. The key point is the construction of several databases of cleaned peakmaps, each covering a given frequency range and built from chunks of data of given duration. All the steps inside the box are repeated for each dataset, belonging to the same or different detectors. Coincidence of candidates found in the various datasets are done in order to reduce the false alarm probability and the surviving ones are followed-up in order to increase their detection significance or to discard them. See text for more details.}\label{fig:scheme}
\end{figure*}

Let us indicate with $t_{\rm coh}$ the signal coherence time, such that over a given observation time the signal frequency varies by at most $1/t_{\rm coh}$.
For a given detector, a collection of {\it Band Sampled Data} (BSD) is built \cite{ref:ornella} from calibrated data, covering the frequency range and the observation time window to be analyzed. This is a very flexible data framework, which we use to perform computationally efficient analyses, as will be clear in the following. An overall picture of the analysis scheme is shown in  Fig.\ref{fig:scheme}. Starting from the BSD data, the blocks inside the figure box correspond to the analysis steps applied to each detector data set. The first step, detailed in \ref{sec:data_base}, consists in the construction of various FFT databases, each from data chunks of a given duration, in order to cover a range of possible signal coherence times (see 1. in Fig.\ref{fig:scheme}), and the selection of the corresponding most significant time-frequency peaks. Next, see \ref{sec:anasche}, for each set of peaks a grid in the sky is built (2.), and for each sky position the peaks are properly shifted in order to correct the Doppler effect due to the detector motion (3.). Candidates, that is potentially interesting points in the signal parameter space, are selected on the frequency histogram of the corrected peaks (4.), and are clustered together according to their characteristics (5.). Once these operations have been done for all the data sets at disposal, a coincidence step is used in order to reduce the false alarm probability (6.), followed by a deeper analysis of the most significant surviving candidates (7.).
Most of the above steps are adapted from the CW all-sky search pipeline described in \cite{ref:fhmethod} and used in \cite{ref:allsky1}, \cite{ref:allsky2}.

\subsection{The peakmap data base}
\label{sec:data_base}
Starting from the ensemble of BSD files, the first step is the construction of a collection of the most significant peaks selected in equalized power spectra, estimated from cleaned data chunks of a given time duration $T_{FFT}$ \cite{ref:pia_sfdb}. The peak frequencies, together with the central time of the corresponding data chunk from which the spectrum has been built, form a time-frequency map called {\it peakmap}. This step corresponds to the first block in Fig.\ref{fig:scheme}.
On a standard search for CW signals emitted by isolated neutron stars, with no prior information on the rotational parameters and on the sky
position of the source, the main constraint on $T_{FFT}$ comes from the Doppler effect due to the relative motion of the source and the detector
\cite{ref:pia_sfdb}.
The value of $T_{FFT}$ affects the pipeline sensitivity, discussed in Sec. \ref{sec:sensitivity}. 

\begin{figure}[htbp]
\includegraphics[width=9cm]{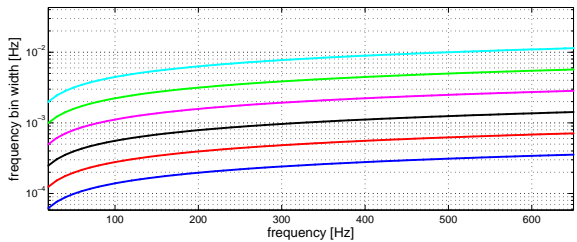} 
\caption{Frequency bin width $\delta f$ (computed at steps of 10Hz) for each of the six DBs, going from DBs built with shorter duration FFTs (top curve) to longer (bottom curve). The FFT time duration, for a given 10 Hz band is simply the inverse of the bin width: $T_{FFT}=1/\delta f$.}\label{fig:BIN_TFFT}
\end{figure}

Since the search described in this paper is aimed at CW signals whose frequency may change randomly, as it will be discussed in Sec. \ref{sec:sensitivity}, the FFT time duration should be chosen accordingly to the signal coherence time. As this quantity in general is not known in advance, we organize the data in a number $N$ of data bases (DB) of cleaned peakmaps $P_{jk}$ each covering a frequency band of 10Hz, where the first index identifies the $j-th$ DB, $DB_j$, with $j=1,...N$, and $k$ ranges over the 10 Hz sub-bands in which the full frequency range to be searched is divided\footnote{For our standard search setup, which covers the range between 20 Hz and 630 Hz, $k=1,..61$.}.
The first DB, $DB_1$, is optimized for sources emitting a CW signal whose frequency is almost constant after the removal of the Earth Doppler effect, as in the case of standard CW searches \cite{ref:allsky1}. As a consequence, the FFT time duration used for each of the peakmaps $P_{1k}$, $T_{FFT,1k}$, is only constrained by the maximum frequency variation due to the Doppler effect and is given, in seconds, by \cite{ref:sergio_cp}
\begin{equation}
T_{FFT,1k} \simeq \frac{1.1\cdot 10^5}{\sqrt{f_{k,{\rm max}}/1 Hz}}, 
\label{TFFT}
\end{equation}
where $f_{k,{\rm max}}$ is the maximum frequency in the $kth$ band.
The next DBs (identified by an index $j>1$) are chosen to detect signals which
 frequency changes randomly with a coherence time comparable to their FFT duration. In order to have a computationally cheap pipeline in the case of an all-sky search these DBs are built using shorter FFT durations.
More specifically, the FFT duration $T_{FFT,jk}$ (with $j>1$) we use to build the peakmap $P_{jk}$ is 
\begin{equation}
T_{FFT,jk}=\frac{T_{FFT,1k}}{2^{(j-1)}}.
\label{TFFTjk}
\end{equation}
For our standard search setup we use $N=6$, which is a reasonable compromise among covering a relatively large range of possible signal coherence times and keeping the anaysis computational load under control.
In Fig. \ref{fig:BIN_TFFT} the frequency resolution of each DB in the standard setup is shown
as a function of the frequency. The FFT duration $T_{FFT}$ is simply the inverse of the
frequency resolution. For each DB, the range of FFT durations is given in the first column of
Tab. \ref{tab:skypo}. 
Each peakmap DB is cleaned by using optimized techniques of data conditioning to reduce noise disturbances \cite{ref:fhmethod}.
These cleaning procedures work at different levels and are of foundamental importance to preserve the search sensitivity against the noise effects.
The first level of data cleaning operates in the time domain and it is finalized  to the reduction of the noise power level via the identification and the removal of big short duration disturbances \cite{ref:pia_sfdb}. Indeed these glitches increases the noise level in the power spectra potentially hiding weak GW signals in the data.
A further cleaning step works in the frequency domain, directly on the peakmap, and is called ``line persistency veto''.  This veto \cite{ref:fhmethod} consists in projecting the peakmap onto the frequency axis, before Doppler correction, and setting a threshold on the basis of the statistical proprieties of this histogram. All the frequency bins exceeding the threshold are then recorded and removed from the analysis.
This cleaning procedure is crucial to remove the strong influence of persistent spectral noise lines before the candidate selection step. Indeed such disturbances can have a great statistical significance, even after the Doppler correction that mitigates their effect by spreading the power of noise lines over more frequency bins.
The action of line removal done at this level allows to improve the search sensitivity and is preferrable to a-posteriori candidate vetoes, see e.g. \cite{ref:demoff}.
\subsection{Sky grid, Doppler correction and candidate selection}
\label{sec:anasche}
\begin{figure}[htbp]
\includegraphics[width=9cm]{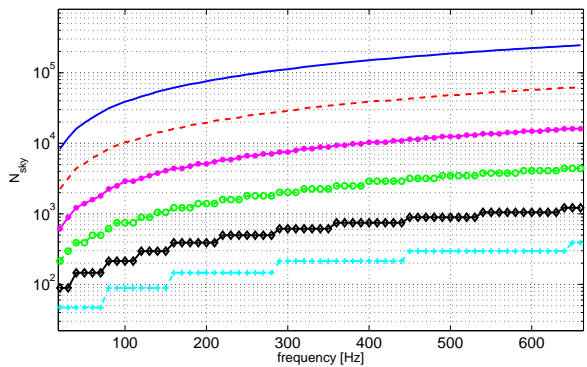} 
\caption{Number of points in the sky grid from 20 Hz to 630 Hz, computed at steps of 10Hz, for each of the six DBs, going from that built with the longest duration FFTs (top curve) to the shortest (bottom curve).}\label{fig:NskyF}
\end{figure}

\begin{figure}[htbp]
\includegraphics[width=9cm]{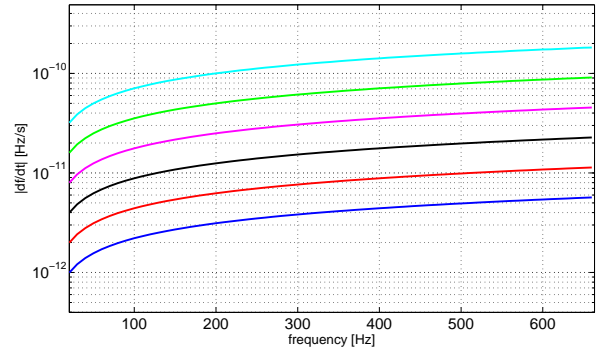} 
\caption{Maximum absolute value of the frequency variation to which the standard search setup would be sensitive to, as a function of the search frequency. The curves correspond to the standard search setup and go from shorter FFT duration (top curve) to longer (bottom curve). One year of observing time is assumed.}\label{fig:SpinDown}
\end{figure}

The next step in the analysis consists in the construction of a proper grid in the sky, such that the frequency variation induced by the Doppler effect moving among two nearby points in the sky is fully contained into one frequency bin. The total number of sky cells depends both on the frequency and on the time length of the FFT \cite{ref:fhmethod} and it is given by
\begin{equation}
N_{sky,jk}=4\pi N^2_{D,jk},
\end{equation}
where
\begin{equation}
N_{D,jk}=\frac{f_{k,{\rm max}}\Omega_{\rm orb}R_{\rm orb}T_{FFT,jk}}{c}, 
\label{Nsky}
\end{equation}
is the maximum number of bins within the band in which the signal frequency is spread by the Doppler effect \cite{ref:fhmethod}. In this equation 
$\Omega_{\rm orb}$ and  $R_{\rm orb}$ are respectively the Earth orbital angular velocity and the
radius of the Earth orbit, and $c$ is the light speed. Figure \ref{fig:NskyF} shows the number
of points in the sky grid for each $DB_{j}$ as a function of the frequency $f_{k,{\rm max}}$. 
The total number of sky points for each of the six DBs of the standard setup is shown in Tab. \ref{tab:skypo}.
\begin{table}[htbp]
\begin{center}
\caption{First column: range of FFT durations, over the frequency band 20-630 Hz, for each of the six DBs of the standard search
setup. Second column: number of sky points for an all-sky search over the band 20-630 Hz using the six standard DBs. Label ``1'' corresponds to the longest duration database, see the top curve in Fig. \ref{fig:NskyF}.}\label{tab:skypo}
\label{tabl2}
\hspace*{-1.0cm}
\begin{tabular}{ccc}
\hline
DB & FFT duration [s] & $N_{sky}$\\ \hline \hline
1&16384-2920&8756163 \\
2&8192-1460&1945429 \\
3&4096-730&510919 \\
4&2048-365&137837 \\
5&1024-182&38195 \\
6&512-91&11940 \\
\hline
\end{tabular}
\end{center}
\end{table} 
Figure \ref{fig:SpinDown} shows the maximum absolute value of the frequency time derivative to which the standard search setup would be sensitive to, i.e. such that the signal frequency over the observation time would remain within one frequency bin. It is clear that, even if the search nominally assumes a null frequency variation (i.e. we do not explicitly search over the frequency derivative $df/dt$), it can detect signals having a non-negligible spin-down or spin-up, up to $\sim 10^{-10}$ Hz/s for DBs built using shorter FFTs. This is connected to the frequency derivative bin given by $\delta \dot{f}=1/(2T_{FFT}T_{\rm obs})$, where $T_{\rm obs}$ is the total observation time, which is larger for shorter FFTs. These values are typically much larger than the spin-up of signals emitted by boson clouds, see Eq. \ref{eq:f_gw_dot}.
For each sky point the Doppler correction for a hypothetical GW source is then performed by properly shifting the time-frequency peaks in order to line-up signal peaks to the same frequency bin. In order to improve the ability of the search to distinguish between true signals and noise disturbances, both the information on the expected signal signature and on the quality of the data is properly used.
Indeed the detector sensitivity is a time function of the source sky position, described by the detector beam-pattern functions and it also depends on the detector noise behavior. All this information is recorded in the peakmap and adaptively used to weight each time-frequency peak \cite{ref:adaphough}. 

For each sky position and for each 1 Hz band the Doppler corrected peakmap is projected onto
the frequency axis and we select the $N_{\rm c}$ bins with the highest count as possible candidates.
The value of $N_{\rm c}$ is choosen as a compromise between the search sensitivity and the computational cost of the next step of the analysis, namely the follow-up. Moreover, candidates are selected by using a procedure that avoids of being blinded by disturbances \cite{ref:fhmethod}. It consists in dividing the frequency band covered by each peakmap into $N_{\rm c}$ sub-bands and selecting the most (or the two most) significant peaks in each sub-band. In this way, even in presence of residual disturbances not removed by the previous cleaning steps, the candidates are uniformly distributed over frequency. For our standard setup a reasonable choice is to select $N_{\rm c}=20$ candidates per 1 Hz band and sky position. The total number of candidates for each detector would be then $N_{\rm tot}\simeq 2.2364 \cdot 10^9$, most of which come, as shown in Tab. \ref{tab:ncand}, from the longest FFT duration DB.
\begin{table}[h!]
\begin{center}
\caption{Total number of candidates for an all-sky search using the six standard DBs, assuming
to select $N_{\rm c}=20$ candidate for each sky position and per 1 Hz band. Label ``1'' corresponds to the longest duration database.}\label{tab:ncand}
\label{tabl2}
\hspace*{-1.0cm}
\begin{tabular}{cc}
\hline
DB &$N_{cand}\cdot 10^9$\\ \hline \hline
1&1.6573 \\
2&0.4261 \\
3&0.1118 \\
4&0.0302 \\
5&0.0084 \\
6&0.0026 \\
\hline
\end{tabular}
\end{center}
\end{table}  
Each candidate is identified by the following set of parameters: sky position, frequency and their related uncertainties, significance, and DB's origin. 
\subsection{Candidate clustering, coincidences and follow-up}
\label{sec:coinc}
A subsequent clustering step aims at reducing the effect of disturbances that tend to create agglomerates of candidates with similar parameters \cite{ref:fhmethod}, \cite{ref:giupeppe}. Trying to gather candidates with a common origin is an additional and efficient ???data cleaning??? step that 
limits the selection of noise candidates. Given the sets of candidates found, following the previous steps, in two or more data sets, belonging to the same or to different detectors, coincidences among their physical parameters (position and frequency) are performed in order to strongly reduce the false alarm probability.
Coincidences are fundamental to reject candidates which are not found, as expected for a real
signal, at nearby points in the parameter space. Given two sets of $M_1$ and $M_2$ candidates, $\mathrm C_1$
and $\mathrm C_2$, and indicating with $C_{1,jl}$ and $C_{2,jm}$ two specific candidates of the 
two sets found in the database $DB_j$, with $l=1,..M_1$ and $m=1,..M_2$, we define their distance as
\begin{equation}
d_{j,lm}=\sqrt{{k_{f;lm}}^2+{k_{\lambda;lm}}^2+{k_{\beta;lm}}^2},
\label{distance}
\end{equation}
being $k_{f;lm}=\frac{f_l-f_m}{\delta_{f}}$,
$k_{\lambda;lm}=\frac{\lambda_l-\lambda_m}{\delta_{\lambda}}$ and
$k_{\beta;lm}=\frac{\beta_l-\beta_m}{\delta_{\beta}}$ the parameter separation in unit of bins
respectively in frequency, ecliptical longitude and ecliptical latitude, with
$\delta_{f},~\delta_{\lambda},~\delta_{\beta}$ the corresponding bin widths defined in
\cite{ref:fhmethod}\footnote{In fact, the frequency bin width has been already defined in this
section and is given by $\delta f=1/T_{FFT}$}. A coincident candidate pair is characterized by
having $d_{j,lm}$ below a given threshold that must be determined on the base of simulations
using software injections of fake signals (a threshold of 3 has been used in the standard CW search described in \cite{ref:allsky0}, \cite{ref:allsky1}). 
If there are more pairs of coincident candidates belonging to the same cluster, only the most significant coincident one is kept.
Coincident candidates are followed-up to reject them or to increase the detection confidence. The best strategy to do this depends on how the signal coherence time is related to the FFT duration used in the analysis. If the signal coherence time is larger than the range of FFT duration used in the initial search, it would be natural to follow the same approach used for standard CW searches, that is using the estimated signal parameters to increase the FFT duration in the semi-coherent step in order to increase the sensitivity, see e.g. \cite{ref:allsky0}. On the other hand, if the signal coherence time is within the range of FFT durations, using longer FFTs would imply a sensitivity loss, see Sec.\ref{sec:sensitivity}, and alternative ways to perform the follow-up are needed. One possibility is that of using algorithms able, in principle, to follow signals with randomly varying frequency, like the Viterbi algorithm \cite{ref:lilli} (which, however, would not be well suited for an all-sky search, due to its computational burden).

\subsection{Computational cost}
\begin{figure}[htbp]
\includegraphics[width=9cm]{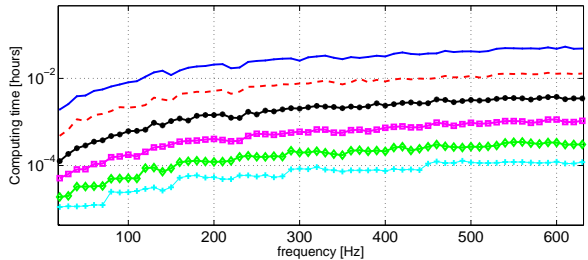} 
\caption{Computing time for the six DBs of the standard setup for each 10 Hz frequency band, assuming one year of data and 2000 cores Xeon E5 - 2620 v3.}\label{fig:CompTime}
\end{figure}

We have estimated the computational cost of the pipeline, up to the candidates selection, by running it on 2 months of simulated Gaussian data and extrapolating performances to a longer observation time. For an all-sky search over the frequency range 20-630 Hz, using the six DBs of the standard setup, the time needed for each DB and for each 10 Hz band, assuming an observation time of one year, is shown in Fig. \ref{fig:CompTime}, assuming the analysis is run on 2000 cores Xeon E5 - 2620 v3. The overall search cost is of about 28 hours. The computational cost can be extrapolated to a search over a wider parameter space considering that it increases roughly as the third power of the maximum search frequency \cite{ref:sergio_cp}, so extending the search up to about 1200 Hz would imply a factor of about eight in the needed computing power, and it is proportional to the number of spin-down or spin-up bins.  
Moreover, at fixed frequency and spin-down range, it scales as the square of the FFT duration.
\subsection{Directed searches} 
\label{sec:direc}
The pipeline can be specialized to the case of a directed search, where the source position is assumed to be known, so that frequency is the only unknown parameter. in a straightforward way. An interesting spot in the sky is represented, for instance, by the Galactic center where a large number of neutron stars and black holes, of the order of several thousands, is expected within a radius of about 1 pc from the center \cite{ref:gc_bh}. Having a restricted region of the sky, we can significantly increase the FFT duration in such a way to improve the search sensitivity {\it if} the signal coherence time is indeed longer than the typical $T_{FFT}$ of the standard all-sky search setup.. An increased FFT duration implies a finer sky resolution and eventually the need to search over a grid of sky points (see Eq. \ref{Nsky}). In Tab. \ref{tab:sky_dir} we show the number of sky points that should be taken into account for a directed search toward the galactic inner parsec for different FFT durations.
\begin{table}[htbp]
\begin{center}
\caption{Number of sky points for a directed search toward a 1 pc region around the Galactic Center for different FFT durations.}\label{tab:sky_dir}
\label{tabl2}
\hspace*{-1.0cm}
\begin{tabular}{cc}
\hline
$T_{FFT}$ [s]&$N_{sky}$\\ \hline \hline
$10^4$&8 \\
$10^5$&800 \\
$10^6$&80000\\
\hline
\end{tabular}
\end{center}
\end{table} 
The computational cost of the directed search strongly depends on the value of $T_{FFT}$ and is estimated to be of the order of 3300 core-days for $T_{FFT}=10^6$ seconds, assuming that a single spin-up(down) bin is taken into account. 

\section{Pipeline sensitivity}
\label{sec:sensitivity}
In this section we derive a theoretical sensitivity estimation for the analysi method previously described. The starting point is the sensitivity of a standard CW search, which is shown in Eq.(61) of \cite{ref:fhmethod}. This expression holds for a semi-coherent search in Gaussian noise in which the residual frequency band over which the signal is spread after Doppler and spin-down corrections, and due to the sidereal modulation, is much smaller than the frequency bin $\delta f=\frac{1}{T_{FFT}}$. If the signal has a wandering frequency over a band $\Delta f$, the sensitivity formula must be corrected by taking into account that if $\Delta f$ is comparable or bigger than $\delta f$, then only a portion of the signal power is confined within a frequency bin. This can be quantified introducing an {\it effective} FFT duration, $T_{FFT,eff}\le T_{FFT}$, related to $T_{FFT}$ by a factor which gives the average fraction of the signal power collected in one frequency bin. The average takes into account the frequency discretization which causes the actual signal frequency to be determined with an uncertainty given by $\pm \delta f/2$. The sensitivity, defined as the minimum signal strain able to produce a candidate in the search, is then a simple generalization of the one valid for standard CW searches, where the FFT duration is replaced by $T_{FFT,eff}$:
\begin{widetext}
\begin{equation}
h_{\rm{0,min}}(t)\approx
\frac{4.02}{N^{1/4}\theta^{1/2}_{\rm thr}}\sqrt{\frac{S_n(f)}{T_{\rm{FFT,eff}}}}\left[\frac{p_0(1-p_0)}{p^2_1}\right]^{1/4}\sqrt{CR_{\rm thr}-\sqrt{2}{\rm erfc}^{-1}(2\Gamma)},
\label{h0min}
\end{equation}
\end{widetext}
where $N={\rm round}\left(\frac{T_{\rm obs}}{T_{FFT}}\right)$, being $T_{\rm obs}$ the observation time, $\theta_{\rm thr}$ is the threshold for peak selection in the equalized spectra, $S_n$ is the detector noise spectral density, $p_0$ is the probability of selecting a peak above the threshold $\theta_{\rm thr}$ if the data contains only noise, $p_1=e^{-\theta_{\rm thr}}-2e^{-2\theta_{\rm thr}}+e^{-3\theta_{\rm thr}}$, $CR_{\rm thr}$ is the threshold on the critical ratio to select candidates on the final projected peakmap, and $\Gamma$ is the chosen confidence level (typically 0.9). The previous formula also takes into account an average over the source sky position and the signal polarization parameters, $cos{\iota}$ and $\psi$. 
\begin{figure*}[htbp]
\includegraphics[width=8cm]{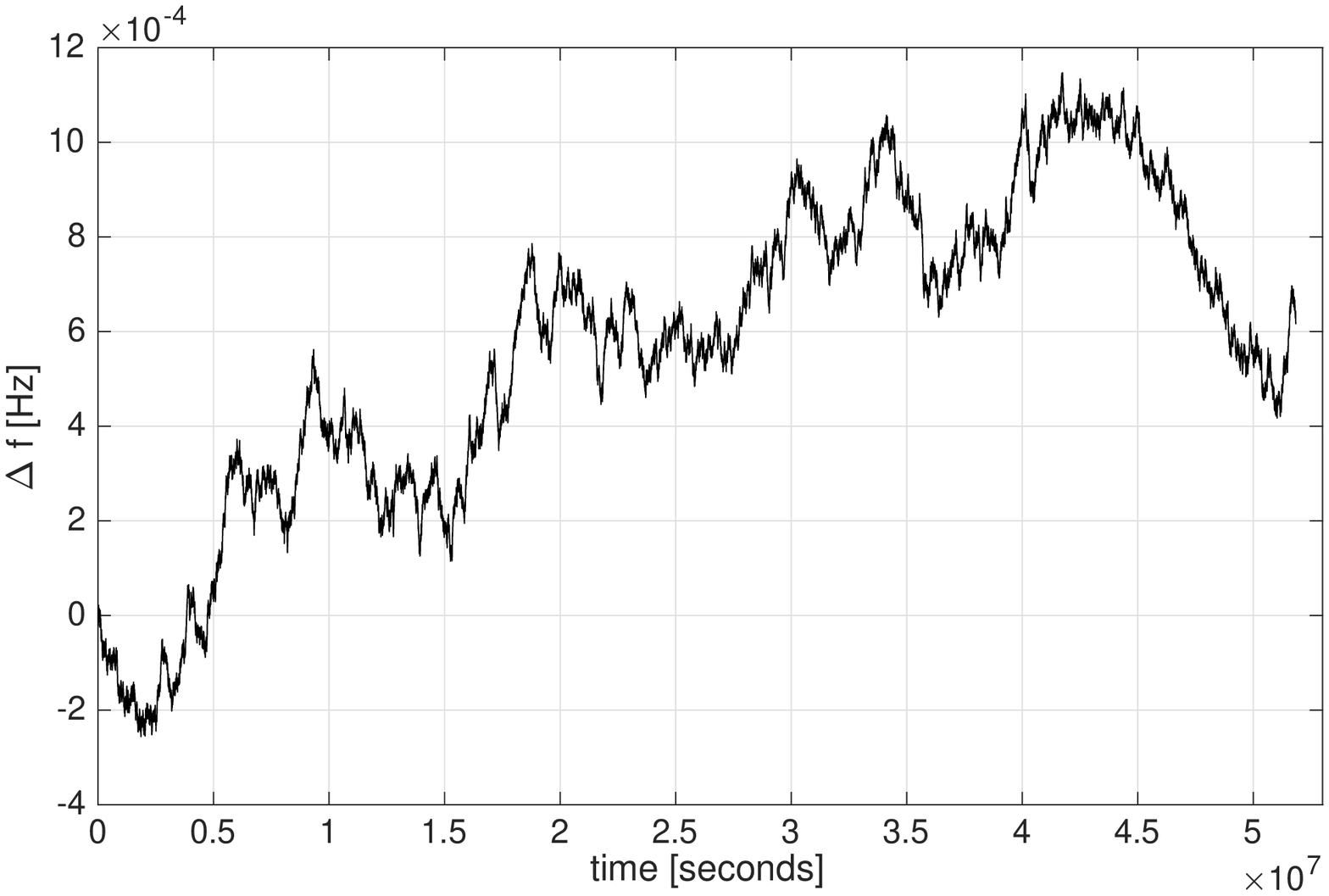}
\includegraphics[width=8cm]{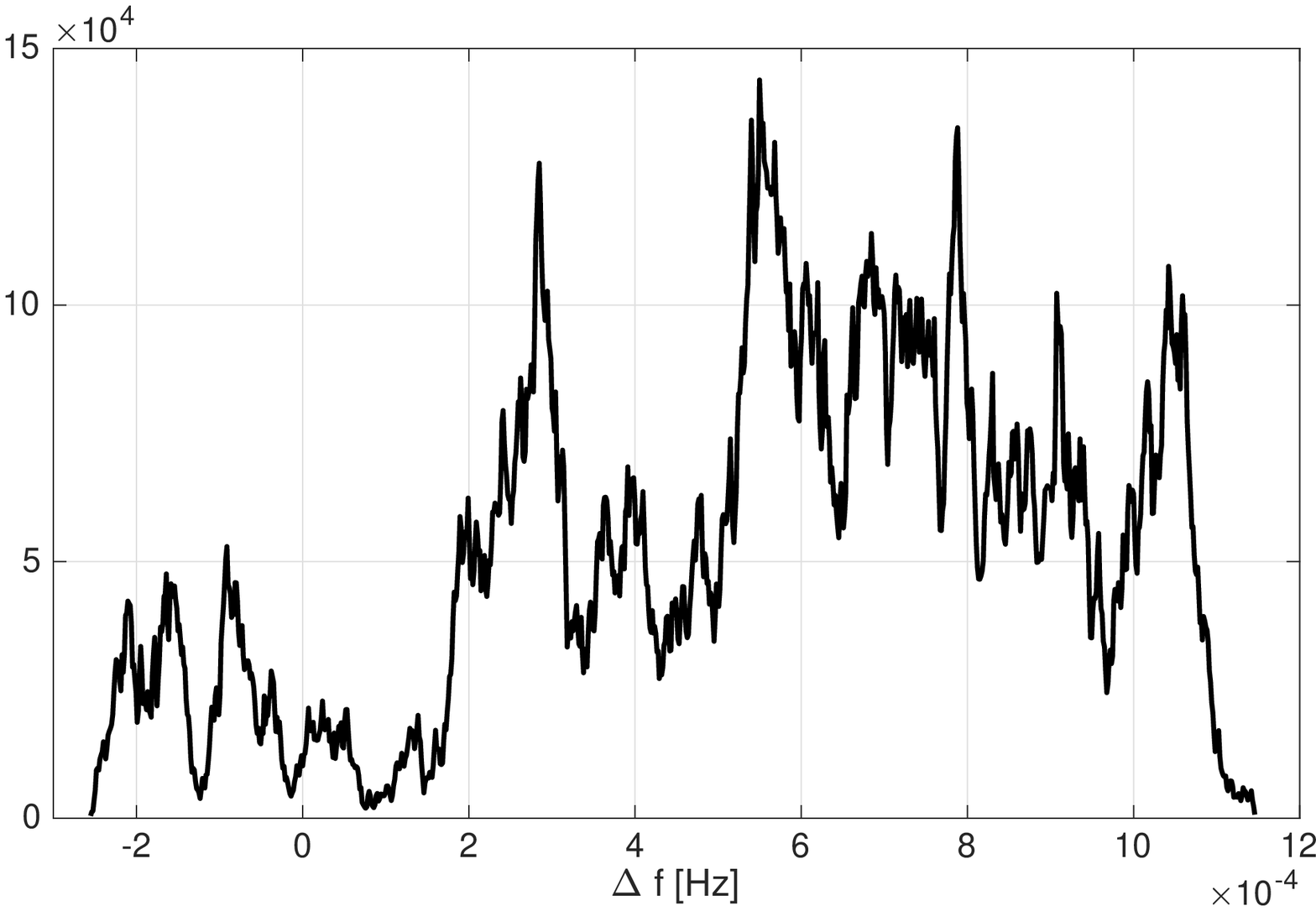}
\caption{Left: an example of wandering frequency, generated with a random walk process, covering a frequency band of 0.0014 Hz. Right: corresponding frequency distribution.}\label{fig:wander_freq}
\end{figure*} 
Figure \ref{fig:wander_freq} shows an example of wandering frequency, generated through a random walk process, over a band $\Delta f=1/713$ Hz =0.0014 Hz, with the corresponding frequency distribution.
From this we have numerically computed the {\it effective} FFT duration as a function of the FFT length, as shown in Fig. \ref{fig:effec_dur}. As expected, $T_{FFT,eff}\simeq T_{FFT}$ for $T_{FFT}\ll 713$ s, as in this case nearly all the signal power remains within one frequency bin. Qualitatively, as $T_{FFT}$ increases, the effective duration stops increasing, and eventually decreases, as a smaller fraction of the signal power is confined in a single frequency bin, with respect to the noise power, before levelling off . The actual quantitative relation between $T_{FFT,eff}$ and $T_{FFT}$ depends on the signal frequency distribution.
\begin{figure}[htbp]
\includegraphics[width=9cm]{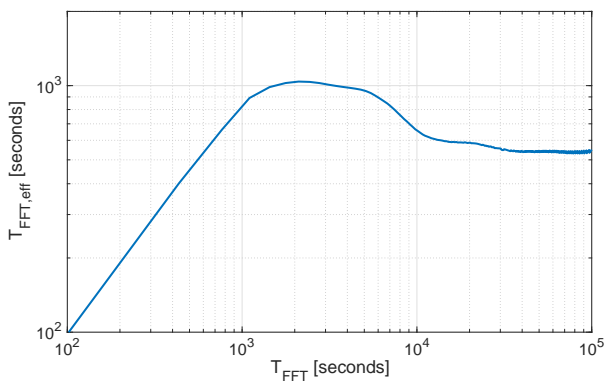} 
\caption{{\it Effective} FFT duration as a function of $T_{FFT}$ for the random walk shown in Fig. \ref{fig:wander_freq}.}\label{fig:effec_dur}
\end{figure} 
By using Eq. \ref{h0min} we can easily compute the sensitivity loss factor for the search of a signal with a given coherence time ($t_{\rm coh}$) as a function of the used FFT duration ($T_{FFT}$). The result, for two signals described by a random walk with, respectively, $t_{\rm coh}=713$ seconds and $t_{\rm coh}=21612$ seconds, is shown in Fig. \ref{fig:sens_loss}, where the sensitivity loss is computed relatively to the best FFT duration. A common feature of the two curves, and a general feature of our method, is that the best sensitivity (the minimum of the two curves in the plot) is reached for an FFT duration $T_{FFT}$ slightly longer than the signal coherence time $t_{\rm coh}$ (in these examples equal, respectively, to 1232 seconds and 25186 seconds). As expected, the sensitivity loss can be significant if there is a big difference between $T_{FFT}$ and $t_{\rm coh}$. 
\begin{figure}[htbp]
\includegraphics[width=9cm]{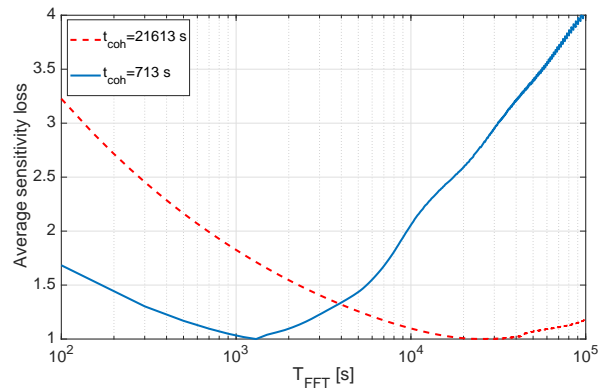} 
\caption{Relative sensitivity loss factor as a function of the FFT duration for signals with coherence times of 713 and 21612 seconds. The best FFT duration corresponds to a sensitivity loss of 1 (i.e. no loss).}\label{fig:sens_loss}
\end{figure}  
On the other hand, the use of the six standard DBs for the all-sky search (as described in Sec. \ref{sec:pipeline}) which differ, each one with respect to the previous and the following, by a factor of two in terms of FFT duration, constraints the sensitivity loss to be less than $\sim 10\%$ for signals which coherence time is within the range of FFT durations covered by the analysis. As already discussed in Sec. \ref{sec:pipeline}, extending the analysis to a wider range of coherence times, which means building more DBs, using shorter and/or longer FFT durations, is just a problem of available computational power and, in any case, could be easily implemented at least for directed searches.
In Fig. \ref{fig:best_sens} we plot the search sensitivity as a function of the frequency, using the O2 LIGO Livingston noise curve of July 20th 2017 (calibration C01; representative of a low noise period), assuming again $t_{\rm coh}=713$ seconds and $t_{\rm coh}=21612$ seconds and that the FFT duration is optimal for those coherence times (corresponding, respectively, to 1232 and 25186 seconds). In the more realistic case, in which there is a mismatch between the FFT duration and the signal coherence time the sensitivity can be corrected by taking into account the loss factor previously described and shown if Fig. \ref{fig:sens_loss}.   
\begin{figure}[htbp]
\includegraphics[width=9cm]{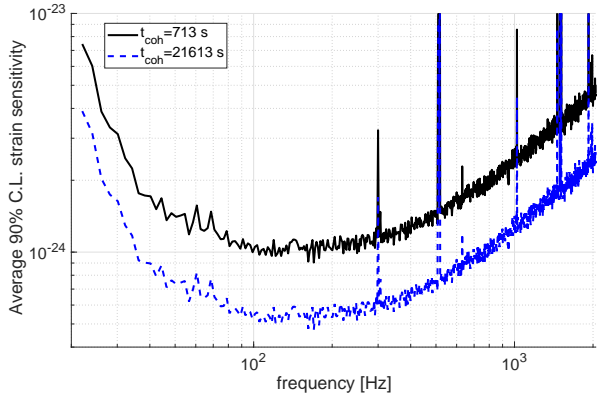} 
\caption{Search strain sensitivity, averaged over sky position and signal polarization parameters, for the LIGO Livingston O2 noise curve of July 20th 2017 (calibration C01). For illustrative purposes, two possible signal coherence times, $t_{\rm coh}=713$ seconds and $t_{\rm coh}=21612$ seconds have been considered and the FFT durations, which maximise the sensitivity for these coherence times, have been used.}\label{fig:best_sens}
\end{figure}  
In the case of boson clouds around black holes, by combining Eqs. \ref{h0approx} and \ref{h0min}, we can determine the search distance reach. Of course, for a given search setup, it depends on several parameters: $M_{\rm bh},~\chi,~m_{\rm b},~t_{cohe}$ and also the age of the emitting system. In order to give some representative results, we have then plotted in Fig. \ref{fig:max_dist} the distance reach as a function of the boson mass $m_{\rm b}$ for various selected black hole masses, taking a constant adimensional initial spin $\chi=0.5$, a fixed signal coherence time $t_{\rm coh}=713$ seconds, assuming to make the search at the optimal FFT duration, $T_{FFT}=1232$ seconds, and using three different ages, $t_{age}=10^3,~10^6,~10^{9}$ years, representative of young, medium age and old systems, respectively. To compute this plot we have taken into account conditions given by Eq. \ref{f_gw_forbid} on the signal frequency, Eq. \ref{eq:mb_forbid} for the duration of the instability phase, and that the GW emission timescale (Eq. \ref{tau_gw}) is not shorter than the instability timescale (Eq. \ref{tau_inst}), i.e. $\tau_{GW}\ge \tau_{\rm inst}$. 
\begin{figure*}[htbp]
\includegraphics[width=12cm]{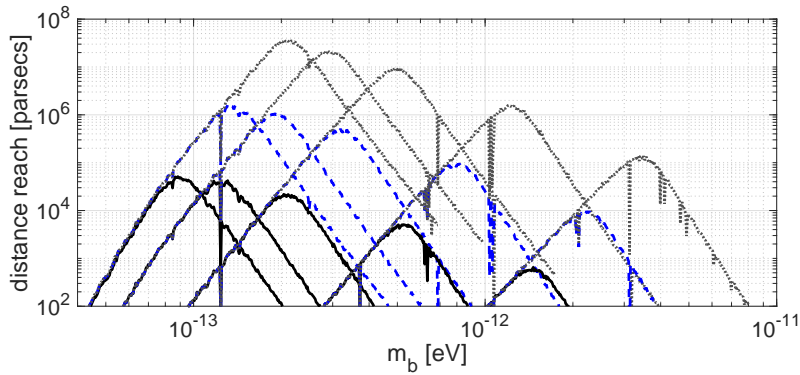} 
\caption{Search distance reach for boson cloud systems of various ages around black holes of different masses. There are seven sets of curves corresponding, going from the right to the left, to black hole masses $M_{\rm bh}=5,~15,~40,~70,~100~M_{\odot}$. For a given black hole mass, the three curves correspond to different ages of the emitting system: $10^3$ years (dotted line), $10^6$ years (dashed line), and $10^9$ years (continuous line). The O2 Livingston sensitivity curve of July 20th 2017 (calibration C01) is used.}\label{fig:max_dist}
\end{figure*}  
This plot shows that boson clouds formed around massive black holes, with mass of the order of 40$M_{\odot}$ or above, could be already detectable by current detectors if they are located within the Galaxy, even in the case of very old systems, provided the signal coherence time is covered in the analysis (and, in general, signals with higher coherence time have better chance of detection). Younger systems, which at fixed parameters emit stronger signals, could be also detectable inside the Galaxy, for lighter black holes, or even outside - up to distances of a few tens of Megaparsecs - for the more massive systems. A mismatch between the signal coherence time and the FFT duration will result in a reduction of the distance reach, depending on the value of $T_{FFT,eff}$. Standard CW searches have been typically done so far into two different regimes concerning the FFT duration: 'quick look' searches have used typical $T_{FFT}$ values of the order of a few thousand seconds \cite{ref:allsky1}; deep 'Einstein@Home' searches have used $T_{FFT}$ up to values of the order of 10 days for restricted frequency ranges \cite{ref:allsky2}. This would imply a sensitivity loss ranging from tens of percent to a factor of several, depending on the actual signal coherence time, if it is significantly different than $T_{FFT}$. Two more considerations are needed here. On one hand, systems forming around more massive black holes, which emit stronger signals, then detectable at larger distances, are characterized by a smaller emission time scale, see Eq. \ref{tau_gw}. This means that, in a given observation period, the probability of catching one or more of these signals is smaller with respect to longer duration signals. Second, more massive black holes are less likely to form, as a consequence of the initial stellar mass function (see e.g. \cite{ref:basti}). In order to estimate the expected number of detections in a given time period, and for a given detector, a simulation of the population of black hole/boson cloud systems would be needed but it is outside the scope of this paper. In fact such kind of study has been already done in \cite{brito1}, under the assumption the signals are perfectly monochromatic and considering an hypothetical full coherent search or, in alternative, a semi-coherent search done combining data segments lasting 250 hours. Those results, based on a rather rough estimation of the analysis sensitivity, should be then taken as somewhat optimistic evaluation of the real detectability as the segment duration cannot be too large due to the computational constraint of all-sky searches. These considerations play also a role in explaining non-detections of signals from boson clouds in the standard CW searches carried on until now and their consequences in terms of constraint on the source properties (Palomba {\it et al.}, paper in preparation). In case of detection, source parameters can be derived. In particular, given an estimation of the signal frequency, from Eq.(\ref{f_gw}), a measure of the boson mass can be done. From Eqs.(\ref{f_gw_forbid}) and (\ref{eq:mb_forbid}), on the other hand, a constraint on the black hole mass can be derived. If a non-zero spin-down is observed, then through Eq.(\ref{eq:f_gw_dot}) the black hole mass could be measured.

\section{Pipeline validation}
\label{sec:efficiency}
\begin{figure}[htbp]
\includegraphics[width=9cm]{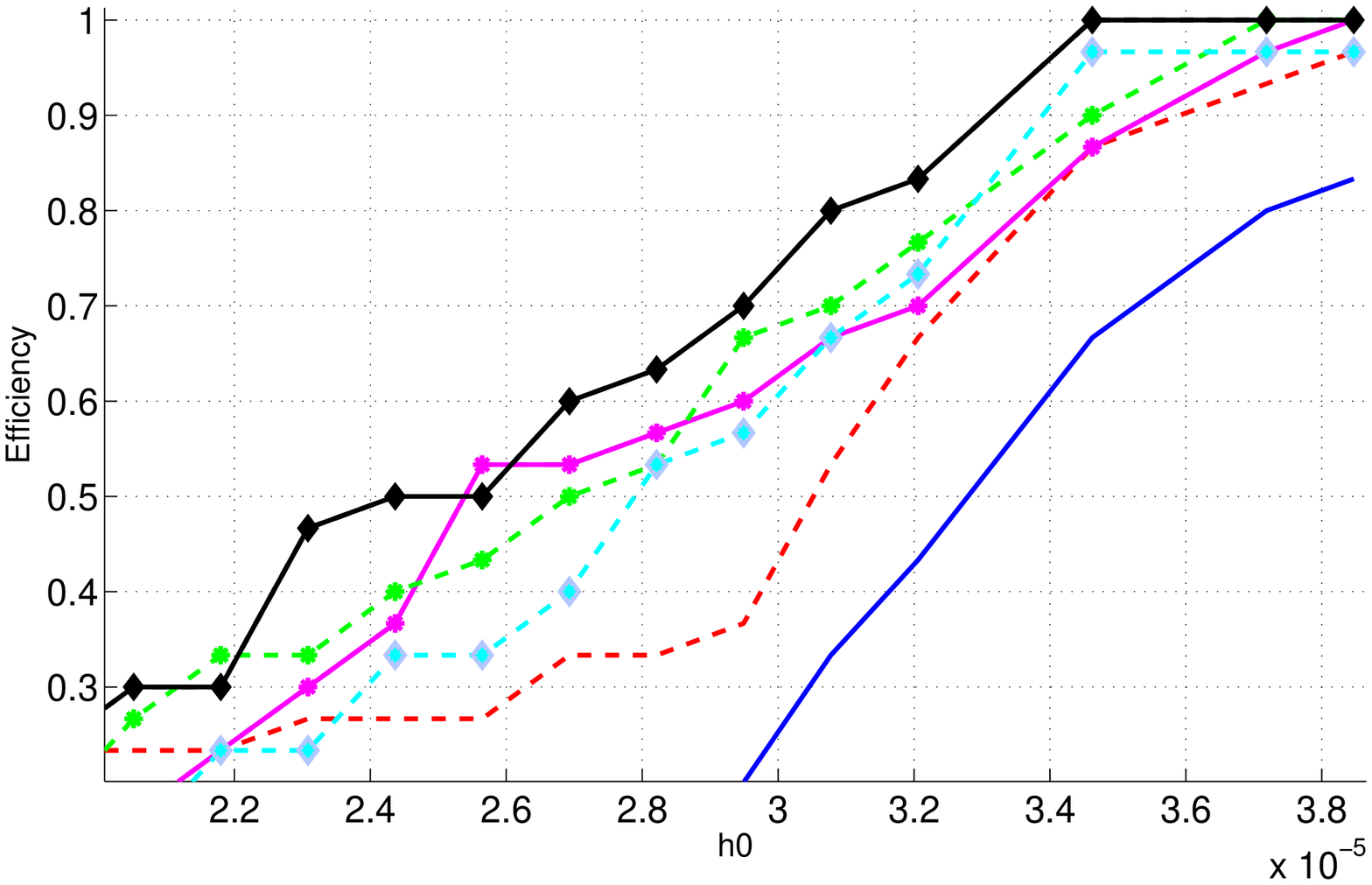} 
\includegraphics[width=9cm]{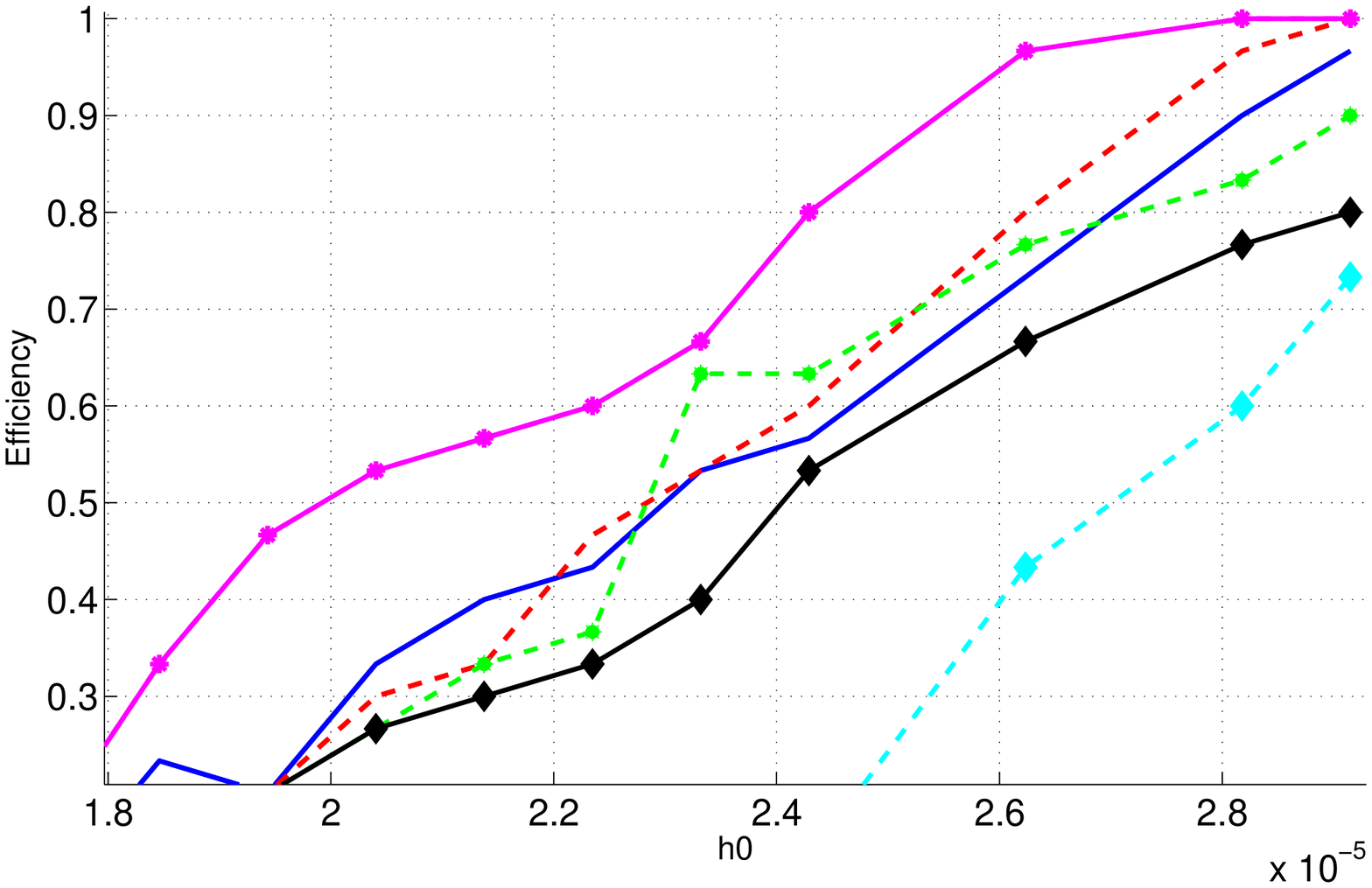} 
\includegraphics[width=9cm]{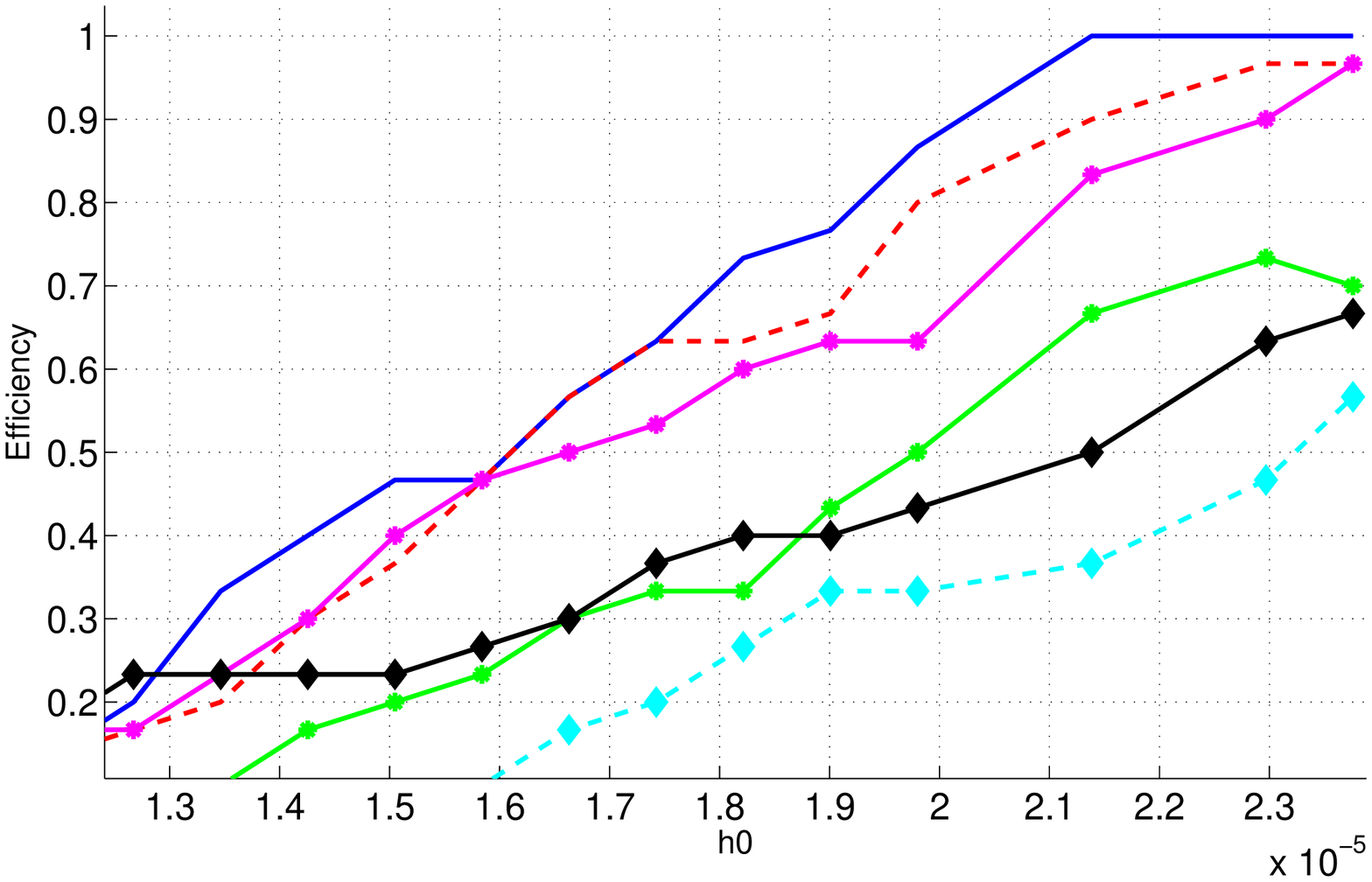} 
\caption{Detection efficiency curves for the six DBs of the standard search setup for gravitational wave signals injected in the band 200-205 Hz and having different coherence times: $t_{\rm coh}=235$ seconds (top),  $t_{\rm coh}=713$ seconds (middle),  $t_{\rm coh}=1615$ seconds (bottom). The FFT durations of each DB, at the injection frequency, are as follows. Continuous line: $T_{FFT}= 5076$ seconds; dotted line: $T_{FFT}=2538$ seconds; asterisk-continuous line: $T_{FFT} = 1268$ seconds; asterisk-dotted line: $T_{FFT}= 634$ seconds ; diamonds-continuous line: $T_{FFT}= 317$ seconds; diamonds-dotted line: $T_{FFT}=158$ seconds. The x-axis is in units of $10^{-20}$.}\label{fig:Eff_235}
\end{figure}

In this section we discuss the pipeline detection efficiency and verify that using a set of FFT databases, covering a range of durations, indeed improves the detectability of signals with an unknown coherence time. 
The capabilities of the pipeline to detect signals have been tested by injecting hundreds of fake signals in simulated 
Gaussian data, covering two months of observing time, with a noise level matching the O2 Ligo Livingston sensitivity 
curve of July 20th 2017 \cite{ref:SensO2}(calibration C01) at a frequency of $~200 Hz$. The choice of using Gaussian noise has been 
dictated by the will to test our pipeline under controlled conditions and, in particular, to understand if it behaves 
as theoretically predicted, without the complication due to non-Gaussian and non-stationary disturbances.
Then, the efficiency plots shown in Fig. \ref{fig:Eff_235} are not affected by non-stationary noise, which instead can pollute real data. Moreover, as this search targets a far broader class of signals with respect to ``standard'' CWs, wandering instrumental lines can highly affect the false alarm rate, therefore a deeper study of the compatibility among signal candidates found in different detectors is of fundamental importance. 
Based on our experience with real data, however, we are confident that the various cleaning steps and veto procedure described 
in Sec. \ref{sec:pipeline}, and already applied in past analyses of LIGO and Virgo data, are robust enough to deal with 
many of the noise outliers that could be found in actual searches. 
Signals simulation and the following analysis have been done as follows. We generated random signal phase evolutions with several different coherence times, from about 100 seconds to about 5000 seconds. For each coherence time we have injected in Gaussian noise a set of 30 signals with the same amplitude, random sky position and polarization parameters, and random frequency in the range between 200 and 205 Hz. These data have been then used to construct a peakmap database on which we performed the full analysis chain up to candidate selection. For every set of injections we have selected $ N_{\rm c}=20$ candidates for each 1 Hz band and sky position, as described in Sec. \ref{sec:pipeline}. A coincidence analysis between the recovered candidate parameters, and the injected signals has been applied to compute how many signals have been detected. We considered a signal as detected if the distance $d$, defined by Eq. \ref{distance}, between the injected signal parameters and the candidate parameters was smaller than a coincidence window $d_{coin}=3$, chosen on the base of the results of the studies performed for the all-sky CW search described in \cite{ref:allsky0}. To gain in computational time, while still maintaining the validity and reliability of the results, for each injected signal we analyzed a limited area of sky centered at the position of the injected signal and such as to cover the maximum allowed coincidence distance $d_{coin}$.
Accidental coincidences have been removed by requiring that the number count $A$ of a candidate coinciding with the injection, should be such to satisfy the following condition:
\begin{equation}
A>2\sigma+{\rm median}(A_{\rm noise})
\label{eq:backgrou}
\end{equation}
being $A_{\rm noise}$ the value of the peakmap projection for the selected candidates in the case of noise only (computed generating a noise data set without injections) and $\sigma$ the standard deviation of the projection. 
For each coherence time the analysis has been repeated for about 15 different signal amplitudes. 

For illustrative purposes, Fig. \ref{fig:Eff_235} shows the experimental efficiency curves for three sets of simulations performed for signal coherence times of 235, 713 and 1615 seconds. Qualitatively similar plots have been obtained for other coherence times. For clarity, the uncertainty on the experimental values, due to the limited number of injections carried out for each signal amplitude, is not shown on the plots and amounts to less than 5$\%$ percent in all cases.
From these efficiency curves we can estimate the search sensitivity at a given confidence level, e.g. 90$\%$, for signals of given coherence time. This number has been obtained considering the ``best'' DB (i.e. characterized by the smallest signal at the chosen confidence level) and making a simple linear interpolation between the two points immediately below and above such level. 

The experimental sensitivity has been compared with the theoretical sensitivity, obtained using Eq.(\ref{h0min}). The comparison is summarized in Tab. \ref{tab:eff_com}. 
The results fully validate the simple idea behind the pipeline design: for each signal with a given amplitude and specific random phase evolution, there is a DB that better than the others is able to collect the signal power.

As can be seen in Fig.\ref{fig:Eff_235}, the best detection efficiency for signals having a coherence time of, respectively, 235, 713 and 1615 seconds, is achieved by DBs built from FFTs with lengths, respectively, of 317, 1268 and 5076 seconds. 
Looking, for instance, at the top plot of the same figure one can note that a signal with a strain amplitude of the order of $3.3\cdot 10^{-25}$, but characterized by a significant frequency variations, of the order of 50 mHz, would have a little chance of being revealed by a standard CW semi-coherent search based on FFTs of duration several thousands of seconds. Indeed for these searches the typical FFT length largely exceeds the optimal one.

Furthermore, the experimental results also validate the theoretical prediction on the sensitivity loss factor shown in Fig. \ref{fig:sens_loss}. As an example, for a signal with coherence time of $713$ seconds and a frequency variation shown in Fig. \ref{fig:wander_freq}, the best performance, corresponding to a loss factor of one, is obtained with $T_{FFT}\simeq 1232$ seconds (see Fig. \ref{fig:sens_loss}). From Fig. \ref{fig:Eff_235} (middle) it can be seen that the best detection efficiency is obtained by the DB corresponding to $T_{FFT}= 1268 s$, that is the FFT duration closer to the optimal one. Moving towards longer (or shorter) $T_{FFT}$ the detection efficiency decreases according to the predictions, within the statistical error (the errors are not plotted in the figures for clarity).
\begin{table}[htbp]
\begin{center}
\caption{Comparison between the theoretical ($h^{\rm th}_{\rm min}$) and the
experimental ($h^{\rm exp}_{\rm min}$) sensitivity, at $90\%$ confidence level, for signals signals characterized by coherence times of $235$, $713$ and $1615$ seconds.
 We report the results of the DB which gave the best performance in each of the considered cases.}
\label{tab:eff_com}
\hspace*{-1.0cm}
\begin{tabular}{ccc}
\hline
 $ t_{\rm coh}$ [s] & $h^{\rm exp}_{\rm min}\cdot 10^{-25} $  & $h^{\rm th}_{\rm
 min}\cdot 10^{-25}$ \\ \hline \hline
$235$& $3.3\pm 0.2$& 3.3\\
$713$&$2.5\pm 0.2 $& 2.6\\
$1615$&$2.0\pm 0.2$&1.9\\
\hline
\end{tabular}
\end{center}
\end{table} 

The beahavior of the different DBs for a signal with a given coherence time can
be also contribute to a better understanding of the signal properties. If, for
instance, a signal is detected with the highest significance with some FFT duration,
and we have still evidence for a signal, but with smaller signal-to-noise ratio,
in both shorter and longer duration DBs, this is a hint we are in presence of a
signal with a wandering frequency and we would be able to constrain its
coherence time. If, on the contrary, the signal is better detected with the
longest duration DB we are considering, then we would not be able to distinguish
among a wandering signal with a longer coherence time and a standard CW signal.
In such situation, a follow-up based on DBs with longer FFT durations could help in
discriminating among the two cases.    
\section{Conclusions}
\label{sec:conclusion}
In this paper we have presented a robust method for all-sky searches of nearly monochromatic
continuous GW signals with a randomly varying frequency. Different physical mechanisms can
produce such kind of signals, like accreting neutron stars in which an equilibrium between
matter accretion and GW emission has been established. A new interesting possibility, currently
subject to rather intense theoretical work, is represented by boson clouds which would
spontaneously form around spinning black hole due to a unstable superradiance process
\cite{arva1}, \cite{brito1}, \cite{arva2}, \cite{brito2}. Altough general, our method is - to
our knowledge - the first serious attempt to deal with such kind of signals from the data analysis point of view. We have tested the pipeline with simulated signals injected in Gaussian data and a full statistical characterization of the method has been derived. The theoretical sensitivity has been computed and translated into a search maximum distance reach. 
The analysis pipeline is computationally cheap and allow to perform all-sky searches of 1 year of data over three detectors, and covering the frequency range between 20 Hz and $\simeq$ 600 Hz, with less than 160,000 core-hours, using an ensemble of six databases of FFTs, with time duration going from a few hundreds seconds to several thousands. By restricting the search to a few potentially interesting spots in the sky, e.g. corresponding to the position of known black hole candidates or to regions where a large black hole population is expected, the FFT duration can be further increased up to values of the order of several days. Code performances, moreover, will probably strongly benefit from being ported on GPU, as we have shown for a somewhat similar analysis problem, namely the computation of the so-called FrequencyHough Transform, for which a speed gain of more than one order of magnitude has been obtained (La Rosa {\it et al.}, in preparation). We plan to apply the peline to data of past LIGO and Virgo science runs in order to test its robustness also in presence of instrumental artifacts. An even more deep insight will be possible, of course, with third generation detectors, like the Einstein Telescope. This opens up a fascinating connection between GW, astrophysics and particle physics. Even in case of non-detection, strong constraints on the population of boson clouds/black hole systems and on their parameters will be established

\section*{Acknowledgement}
We want to thank Paolo Pani and Richard Brito for useful comments on this paper.

\end{document}